\documentclass[twocolumn]{aastex631}

\usepackage{amssymb, amsfonts, amsmath,framed}

\usepackage{bm}
\usepackage{graphicx}
\usepackage{epstopdf,color,verbatim}
\usepackage{threeparttable}
\usepackage{amsmath}
\usepackage{amsmath}	
\usepackage{amssymb}	
\usepackage[]{txfonts}

\expandafter\ifx\csname package@font\endcsname\relax\else
 \expandafter\expandafter
 \expandafter\usepackage
 \expandafter\expandafter
 \expandafter{\csname package@font\endcsname}
\fi
\expandafter\ifx\csname package@font\endcsname\relax\else
 \expandafter\expandafter
 \expandafter\usepackage
 \expandafter\expandafter
 \expandafter{\csname package@font\endcsname}
\fi
\newcommand{\figg}[1]{Fig.~\ref{fig:#1}}
\newcommand{\eqq}[1]{Eq.~(\ref{eq:#1})}
\def\bi{\begin{itemize}}
\def\ei{\end{itemize}}
\def\bqy{\begin{eqnarray}}
\def\eqy{\end{eqnarray}}

\newcommand{\ls}{\textcolor{red}}
\newcommand{\vatot}{v_{\rm{A}}}
\newcommand{\vain}{v_{\rm{A,in}}}
\newcommand{\vaout}{v_{\rm{A,out}}}
\newcommand{\csw}{c_{sw}}

\newcommand{\ktil}{\tilde{k}}

\newcommand{\wtil}{\tilde{\omega}}
\newcommand{\gammaw}{\Gamma_{w}}
\newcommand{\bsh}{v}

\begin{document}
\title{The Kelvin-Helmholtz instability at the boundary of relativistic magnetized jets}

\author{Anthony Chow}
\email{kc3058@columbia.edu}
\affiliation{Department of Astronomy and Columbia Astrophysics Laboratory, Columbia University, New York, NY 10027, USA}
\author{Jordy Davelaar}
\email{jrd2210@columbia.edu}
\affiliation{Department of Astronomy and Columbia Astrophysics Laboratory, Columbia University, New York, NY 10027, USA}
\affiliation{
 Center for Computational Astrophysics, Flatiron Institute, 162 Fifth Avenue, New York, NY 10010, USA}
\author{Michael E. Rowan}
\email{michael.rowan@amd.com}
\affiliation{Advanced Micro Devices, Inc., Santa Clara, CA, USA}
\author{Lorenzo Sironi}
\email{lsironi@astro.columbia.edu}
\affiliation{Department of Astronomy and Columbia Astrophysics Laboratory, Columbia University, New York, NY 10027, USA}

\correspondingauthor{Anthony Chow}

\begin{abstract}
We study the linear stability of a planar interface separating two fluids in relative motion,  focusing on conditions appropriate for the boundaries of relativistic jets. The jet is magnetically dominated, whereas the ambient wind is gas-pressure dominated. We derive the most general form of the dispersion relation and provide an analytical approximation of its solution for an ambient sound speed much smaller than the jet Alfv\'en speed $\vatot$, as appropriate for realistic systems. The stability properties are chiefly determined by the angle $\psi$ between the  wavevector and the jet magnetic field. For $\psi=\pi/2$, magnetic tension plays no role, and our solution resembles the one of a gas-pressure dominated jet. Here, only sub-Alfv\'enic jets are unstable ($0<M_e\equiv(v/\vatot)\cos\theta<1$, where $v$ is the shear velocity and $\theta$ the angle between the velocity and the wavevector). For $\psi=0$, the free energy in the velocity shear needs to overcome the magnetic tension, and only super-Alfv\'enic jets are unstable ($1<M_e<\sqrt{(1+\Gamma_w^2)/[1+(\vatot/c)^2\Gamma_w^2]}$, with $\Gamma_w$ the wind adiabatic index). Our results have important implications for the propagation and emission of relativistic magnetized jets.
\end{abstract}


\section{Introduction}
The Kelvin-Helmholtz instability (KHI) \citep{Helmholtz, Kelvin}--- at the interface of two fluids in relative motion --- is one of the most ubiquitous and well-studied instabilities in the Universe. Since the pioneering works of \citet{chandrasekhar_1961}, the linear theory of the KHI has been investigated under a variety of conditions \citep{turland_scheuer_1976,blandford_pringle_1976,ferrari_trussoni_zaninetti_1980,pu_kivelson_1983, kivelson_zu-yin_1984, bodo_mignone_rosner_2004, osmanov_mignone_massaglia_bodo_ferrari_2008,Blumen_75,Ferrari_78, Sharma_98, 
Prajapati_10,Sobacchi_18, Berlok_19,rowan_phd, Hamlin_Newman_2013, Bodo_Mamtsashvili_Rossi_2013, Bodo_Mamtsashvili_Rossi_2016,Bodo_Mamtsashvili_Rossi_2019, Pimentel_Lora-Clavijo_2019}, depending on whether the relative motion is non-relativistic or ultra-relativistic, whether the two fluids have comparable or different properties (respectively, ``symmetric'' or ``asymmetric'' configuration), whether the flow is incompressible or compressible, and whether or not the fluids are magnetized. 

The boundaries of relativistic astrophysical jets may be prone to the KHI, given the relative (generally, ultra-relativistic) shear velocity between the jet and the ambient medium (hereafter, the ``wind''). In jet boundaries with flow-aligned magnetic fields, KH vortices can wrap up the field lines onto themselves, leading to particle acceleration via reconnection \citep{rowan_phd,sironi_21}. Particles pre-energized by reconnection \citep[e.g.,][]{ss_14,zhang_sironi_21,sironi_22} can then experience shear-driven acceleration \citep{rieger_19, reville_21,wang_22} --- i.e., particles  scatter in between regions that move toward each other because of the velocity shear, akin to the Fermi process in converging flows \citep{fermi_49}. The KHI may then constitute a fundamental building block for our understanding of the origin of radio-emitting electrons in limb-brightened relativistic jets (e.g., in Cygnus A \citep{boccardi_16} and M87 \citep{walker_18}), and for the prospects of shear-driven acceleration  at jet boundaries in generating Ultra High Energy Cosmic Rays.

A study of the KHI in this context needs to account for the unique properties of the boundaries of relativistic jets. First, the relative motion between the jet and the wind can be ultra-relativistic; second, while the wind is likely gas-pressure dominated, relativistic jets are believed to be magnetically dominated \citep{blandford_77}, i.e., an asymmetric configuration. The linear stability properties of the KHI in this regime (of relativistic, asymmetric, magnetized flows) are still unexplored. In this {\it Letter}, we  derive the most general form of the  dispersion relation for the KHI at the interface between a magnetized relativistic jet and a gas-pressure-dominated wind. We also provide an analytical approximation of its solution for wind sound speeds much smaller than the jet Alfv\'en speed, as appropriate for realistic astrophysical systems.
\section{Setup}
\begin{figure}[!h]
	\centering
	\includegraphics[width=0.5\textwidth]{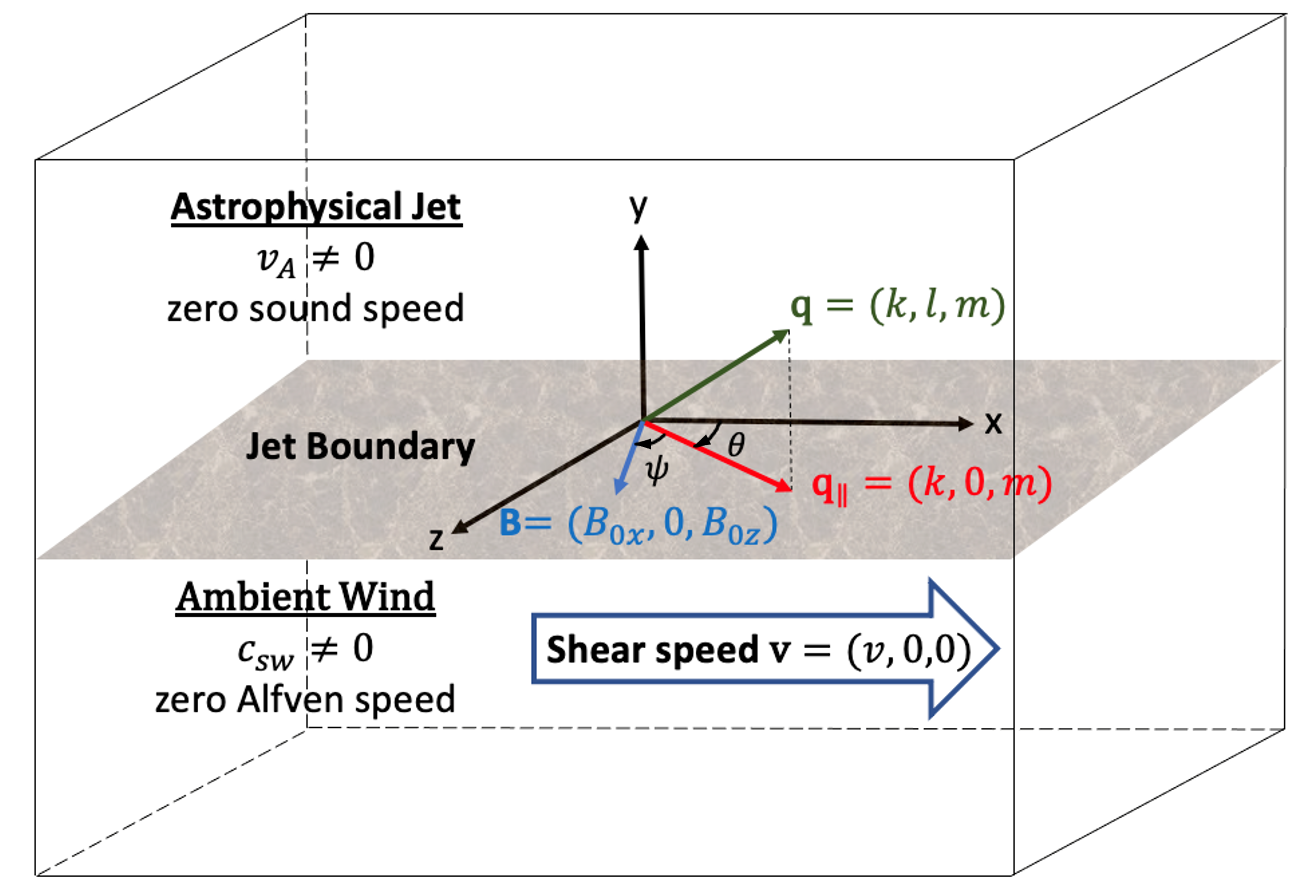}
\caption{A 3D schematic diagram of the boundary of the relativistic jet. The boundary (grey color) is located in the $x-z$ plane. Above and below the boundary are the magnetically-dominated cold jet and the unmagnetized gas-pressure-supported ambient wind, respectively. $\mathbf{q_\parallel}$ is the projection of the wavevector $\mathbf{q}$ onto the boundary. The jet is at rest and the wind has a relative shear speed of $\mathbf{v}$. The magnetic field in the jet is $\mathbf{B}$. $\theta$ is the angle between $\mathbf{q_\parallel}$ and $\mathbf{v}$ while $\psi$ is the angle between $\mathbf{B}$ and $\mathbf{q_\parallel}$.}
\label{fig:schematic_setup}
\end{figure}

We consider a planar vortex-sheet interface in the $x$--$z$ plane at $y=0$, as shown in Fig.~\ref{fig:schematic_setup}. The jet ($y>0$) is cold and magnetized, with field $\mathbf{B}_{0j}=(B_{0x},0,B_{0z})$ lying in the $x$--$z$ plane, and Alfv\'en speed $\vatot$. The ambient wind ($y<0$) is gas-pressure supported (with sound speed $c_{sw}$) and has a vanishing magnetic field. We use the subscript ``\textit{j}'' for the jet and ``\textit{w}'' for the wind. We solve the system in the jet rest frame, where the wind moves with velocity $\mathbf{v}=v\,\hat{x}$. We adopt Gaussian units such that $c=4\pi=1$ and define all velocities in unit of $c$.

We describe the flow with the equations of relativistic magnetohydrodynamics (RMHD) \citep[e.g.,][]{mignone_mattia_bodo_2018,rowan_phd}:
\begin{subequations}\label{eq:RMHD_equations}
\begin{align}
    & \frac{\partial (\rho \gamma)}{\partial t}+\nabla \cdot (\rho \gamma \mathbf{v})=0 \\
    &\frac{\partial}{\partial t} (w\gamma^2\mathbf{v}) + \nabla\cdot(w\gamma^2\mathbf{vv})+\nabla p = \rho_e \mathbf{E}+\mathbf{J}\times \mathbf{B} \label{eq:momentum_eqn}\\
    & \frac{\partial \mathbf{B}}{\partial t} + \nabla\times\mathbf{E}=0 \label{eq:faraday_eqn}\\
    & \frac{\partial \mathbf{E}}{\partial t}-\nabla \times\mathbf{B}=-\mathbf{J} \label{eq:ampere_eqn}\\
    & \frac{\partial}{\partial t}(w\gamma^2-p)+\nabla\cdot (w \gamma^2 \mathbf{v})=\mathbf{J}\cdot\mathbf{E}
\end{align}
\end{subequations}
supplemented with the divergence constraints
\begin{align}
    \nabla \cdot \mathbf{E}=\rho_e, \quad \nabla \cdot \mathbf{B}=0
\end{align}
Here, $\rho, \rho_e, \mathbf{J}, \mathbf{v}, \gamma, \mathbf{B},\mathbf{E}, w$ and $p$ are the rest-mass density, charge density, current density, fluid velocity, Lorentz factor ($\gamma=1/\sqrt{1-v^2}$), magnetic field, electric field, gas enthalpy density and pressure, respectively.
For an ideal gas with adiabatic index $\Gamma$, the enthalpy can be written as $w=\rho+ \Gamma p/(\Gamma-1)$.

We assume a cold and magnetically-dominated jet, with Alfv\'en speed $\vatot^2=\vain^2+\vaout^2$, where
\begin{align}
    \vain\!=\!\sqrt{\frac{B_{0x}^2}{w_{0j}\!+\!B_{0x}^2\!+\!B_{0z}^2}}, \text{  } \vaout\!=\!\sqrt{\frac{B_{0z}^2}{w_{0j}\!+\!B_{0x}^2\!+\!B_{0z}^2}}
\end{align}
and the jet enthalpy density is $w_{0j}\approx \rho_{0j}$ for a cold jet. The wind has negligible magnetic field and is gas-pressure supported, with sound speed \citep{mignone_mattia_bodo_2018}
\begin{align}
    c_{sw}=\sqrt{\frac{w_{0w}-\rho_{0w}(\partial w_{0w}/\partial \rho_{0w})} {(\partial w_{0w}/\partial p_{0w})-1}\frac{1}{w_{0w}}}=\sqrt{\Gamma_w\frac{p_{0w}}{w_{0w}}}
\end{align}
where $w_{0w}$ is the wind enthalpy density. From pressure balance across the interface,
\begin{align}\label{eq:pressure_balance}
    \frac{1}{2}(B_{0x}^2+B_{0z}^2)=\frac{c_{sw}^2 w_{0w}}{\Gamma_w} \Rightarrow \frac{w_{0w}}{w_{0j}}=\frac{1}{2} \frac{\vatot^2\Gamma_{w}}{(1-\vatot^2)c_{sw}^2}~,
\end{align}
where $\Gamma_{w}$ is the wind adiabatic index.


\section{Dispersion relation}
The dispersion relation of surface waves at the interface can be found from the dispersion relations of body waves in both the jet and the ambient wind, together with the displacement matching at the interface. The dispersion relations of body waves in each of the two fluids can be found by linearizing Eqs.~(\ref{eq:RMHD_equations}), such that the perturbed variables take the form $\varphi\approx\varphi_0+\varphi_1$, where $\varphi_0$ and $\varphi_1$ are the background and the first-order perturbed variables respectively. The perturbed electric field in the jet is $\mathbf{E}_1=-\mathbf{v}_1\times \mathbf{B}_{0j}$ in the ideal MHD limit \footnote{Resistive effects are likely important in the non-linear stages \citep{sironi_21}, but not for the linear analysis presented here.}, where $\mathbf{v}_1$ is the perturbed velocity in the jet frame.

In the jet, we consider perturbed variables $\varphi_1$ of the form $\varphi_1 \propto e^{i(\mathbf{q}\cdot \mathbf{x}-\omega t)}$ where $\mathbf{q}=(k,l_j,m)$ is the complex wavevector and $\omega$ is the complex angular frequency,
both defined in the jet rest frame. Note that $\rm{Im}(\omega)>0$ implies that the amplitude of the wave grows exponentially, i.e., an instability. We define the angle $\theta$ between the projection of the wavevector onto the $x$-$z$ plane and the direction $\hat{x}$ of the shear flow velocity such that \begin{equation}
 \cos\theta = \frac{k}{\sqrt{k^2+m^2}}~.
\end{equation}
Similarly, we define the angle $\psi$ between the wavevector projection onto the $x-z$ plane and the jet magnetic field such that
\begin{align}
    \cos\psi = \frac{k\vain+m\vaout}{\vatot\sqrt{k^2+m^2}}~.
\end{align}
For a magnetized cold jet, the dispersion relation of  its body waves describes magnetosonic waves in the cold plasma limit:
\begin{align}
    \label{eq:jet_DR_lab_frame}
    &\omega[\omega^2-(k \vain + m \vaout)^2]\nonumber \\
    & \quad \quad [\omega^2-(k^2+l_j^2+m^2)\vatot^2]=0~.
\end{align}
In the wind, we consider perturbed variables $\varphi_1$ of the form $\varphi_1 \propto e^{i(\mathbf{\tilde{q}}\cdot \mathbf{x}-\tilde{\omega} t)}$ where $\mathbf{\tilde{q}}=(\tilde{k},l_w,m)$ is the complex wavevector and $\tilde{\omega}$ is the complex angular frequency,
both defined in the wind rest frame. For an unmagnetized wind, the dispersion relation of its body waves reduces to the one of sound waves, $\wtil^2-(\ktil^2+l_w^2+m^2)\csw^2=0$. By Lorentz transformations of $\wtil=\gamma(\omega - k v)$ and $\ktil=\gamma(k-v \omega)$, we obtain
\begin{align} \label{eq:wind_DR_lab_frame}
    \gamma^2(\omega-k v)^2= c_{sw}^2[l_w^2+m^2+\gamma^2(k-\omega v)^2]~.
\end{align}

Since $l_j$ and $l_w$ are Lorentz invariant, by solving Eq.~(\ref{eq:jet_DR_lab_frame}) and  Eq.~(\ref{eq:wind_DR_lab_frame}) for $l_j$ and $l_w$ respectively, we can construct a Lorentz invariant ratio:
\begin{align}\label{eq:lpluslminus2}
\frac{l_w^2}{l_j^2}= \frac{\vatot^2 [\gamma^2(\omega-k v)^2-c_{sw}^2(m^2+\gamma^2(k-\omega v)^2)]}{c_{sw}^2[\omega^2-(k^2+m^2)\vatot^2]}
\end{align}
An independent way of obtaining $l_w/l_j$ is to simultaneously solve the linearized RMHD equations, Eqs.~(\ref{eq:RMHD_equations}), together with the first order pressure balance equation
\begin{align}
    B_{0x}B_{1x}+B_{0z}B_{1z} = p_{1w}
\end{align}
and the displacement matching condition at the interface
\begin{align}
    \frac{v_{1y,j}}{\omega}=\frac{v_{1y,w}}{\gamma(\omega-kv)}~,
\end{align}
yielding
\begin{align} \label{eq:lpluslminus}
    \frac{l_w}{l_j}=\frac{\gamma^2(1-\vatot^2)(\omega-kv)^2}{\omega^2-(k\vain+m\vaout)^2} \frac{w_{0w}}{w_{0j}}~.
\end{align}
Using Eq.~(\ref{eq:pressure_balance}), we can eliminate $w_{0w}/w_{0j}$ from Eq.~(\ref{eq:lpluslminus}) and  finally, the dispersion relation for the surface wave at the interface can be obtained by equating Eq.~(\ref{eq:lpluslminus2}) and the square of Eq.~(\ref{eq:lpluslminus}):
\begin{align} \label{eq:interface_DR_raw}
    &\frac{\gamma^2(\omega-k\bsh)^2-c_{sw}^2(m^2+\gamma^2(k-\omega\bsh)^2)}{\omega^2-(k^2+m^2)\vatot^2} \nonumber \\
    & \quad \quad =\frac{1}{4} \frac{\vatot^2\gamma^4(\omega-kv)^4 \Gamma_w^2}{[\omega^2-(k\vain+m\vaout)^2]^2 c_{sw}^2} ~.
\end{align}
By introducing the following notations, 
\begin{align} \label{eq:parameters}
    &\phi = \frac{\omega}{\vatot \sqrt{k^2+m^2}}, \quad M=\frac{v}{\vatot}, \quad
     \epsilon=\frac{c_{sw}}{\vatot}~,
\end{align}
Eq.~(\ref{eq:interface_DR_raw}) can be rewritten as \citep{Sobacchi_18,rowan_phd}
\begin{align} \label{eq:interface_DR_param}
    &4 \epsilon^2 (1-M^2\vatot^2)(\cos^2\psi-\phi^2)^2 \nonumber \\  & \quad [\epsilon^2(1-2M\vatot^2\phi\cos\theta +M^2\vatot^2(\cos^2\theta-1+\vatot^2\phi^2)) \nonumber \\
    & \quad -(M\cos\theta-\phi)^2] =(M\cos\theta-\phi)^4(1-\phi^2)\Gamma_w^2
\end{align}
The dispersion relation in Eq.~(\ref{eq:interface_DR_param})  holds for arbitrary values of $c_{sw}$, $\vatot$, $v$, $\cos\theta$ and $\cos\psi$, subject only to the assumptions of a cold jet and an unmagnetized wind.

Since Eq.~(\ref{eq:interface_DR_param}) is a sextic equation in $\phi$, it has a total of six (generally, complex) roots. However, not all of them may be acceptable. First, not all of the solutions will satisfy \eqq{lpluslminus}, since we have introduced spurious roots  when squaring it. Also, by the Sommerfeld radiation condition \citep{sommerfeld_1912}, only outgoing waves should be retained. This requires $\text{Im}(l_w) < 0$ and $\text{Im}(l_j)>0$. The expressions for $l_w$ and $l_y$ can be obtained from the derivation of Eq.~(\ref{eq:lpluslminus}), so the Sommerfeld condition can be expressed as
\begin{subequations}
\begin{eqnarray}
\text{Im}(l_w) & = & \text{Im}\left ( \frac{(\phi-M\cos\theta)^2}{\phi} \right) <0 \\
\text{Im}(l_j) & = & \text{Im}\left ( \frac{\phi^2- \cos^2\psi}{\phi}\right) >0
\end{eqnarray}
\end{subequations}

\section{Analytical approximation}
Since in general a sextic equation has no algebraic roots \citep{abel_1826}, only approximate solutions of $\phi$ in Eq.~(\ref{eq:interface_DR_param}) can be obtained. We first note that the parameters in Eq.~(\ref{eq:parameters}) are chosen such that for a realistic wind with $c_{sw}\ll \vatot$, we have $\epsilon \ll 1$, whereas the other parameters do not depend on $c_{sw}$. We then expand $\phi$ as a power series of $\epsilon$ of the form
    $\phi \approx c_0 + c_1 \epsilon + c_2\epsilon^2$,
where $c_0, c_1$ and $c_2$ are constant with respect to $\epsilon$ and terms higher than $\epsilon^2$ are ignored. Substituting this into Eq.~(\ref{eq:interface_DR_param}) and comparing coefficients of various powers of $\epsilon$ on both sides, we can find an approximate solution for all six roots of Eq.~(\ref{eq:interface_DR_param}). If we define an effective Mach number
\begin{align}
    M_e \equiv M\cos\theta=(v/v_{\rm A})\cos\theta~,
\end{align}
$\mu\equiv \cos^2\psi-M_e^2$, and recognize that $\gamma^{-2}=1-M^2\vatot^2$, {then the approximate roots that correspond to the unstable modes can be written as
\begin{align}
    &\phi_{(M_e<1)}=M_e + \Lambda_+ \epsilon - \Sigma_+ \epsilon^2 ~,\label{eq:approx1}\\
    &\phi_{(M_e>1)}=M_e - \Lambda_- \epsilon + \Sigma_- \epsilon^2~, \label{eq:approx2}
\end{align}
}

where
\begin{eqnarray}
    \Lambda_\pm &= &\sqrt{-\frac{2\,(\mu^2 \pm \lambda)}{\gamma^2(1-M_e^2)\Gamma_w^2}}~,  \label{eq:first_order_terms} \\
    \lambda &= &\sqrt{\mu^4+\mu^2(1-M_e^2)(1-M_e^2\vatot^2)\Gamma_w^2}~. \label{eq:approx_phi_lambda}
\end{eqnarray}
We find that the first order term $\Lambda_{\pm}\epsilon$ generally provides a good approximation of the numerical solution for $\phi$. However, the second order term (which we write explicitly in Appendix \ref{app2}) is
required for identifying the physical solutions that satisfy  \eqq{lpluslminus} and the Sommerfeld condition. At zeroth order in $\epsilon$, the real part of the solution (i.e., the phase speed of unstable modes) is $\phi=M_e$, or equivalently $\omega/k=v$, i.e., unstable modes are purely growing in the wind frame.

\begin{figure}[!h]
	\centering
	\includegraphics[width=0.5\textwidth]{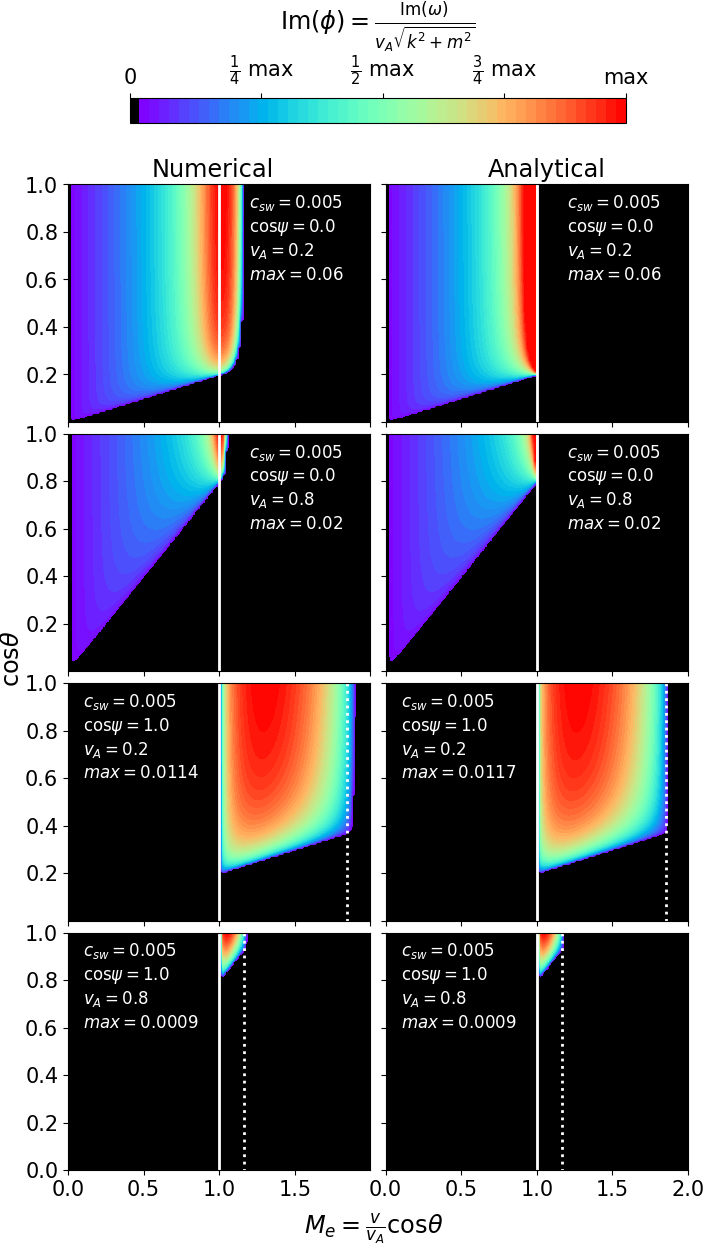}
	\caption{Dependence of the instability growth rate $\text{Im}(\phi)$ on $\theta$ and $M_e$, for two choices of $\vatot$ and two choices of $\cos\psi$, as indicated in the plots.
	The left and right columns represent the numerical and analytical solutions, respectively.  For $\cos \psi=0$, the maximum growth rate of the analytical solution is capped at its numerical counterpart to avoid the divergence at $M_e=1$. In all the panels, $\text{Im}(\phi)$ is then normalized to its maximum value, which is quoted in the panels themselves. The vertical dotted lines show the analytical upper bound on $M_e$ when $\cos\psi=1$, see \eqq{bounds_for_psi_1}. The vertical solid white lines indicate $M_e=1$.
	}
	\label{fig:numerical_vs_analytical_costheta}
\end{figure}

\begin{figure}[!h]\label{fig:numerical_vs_analytical_spanning_psi}
	\centering
	\includegraphics[width=0.5\textwidth]{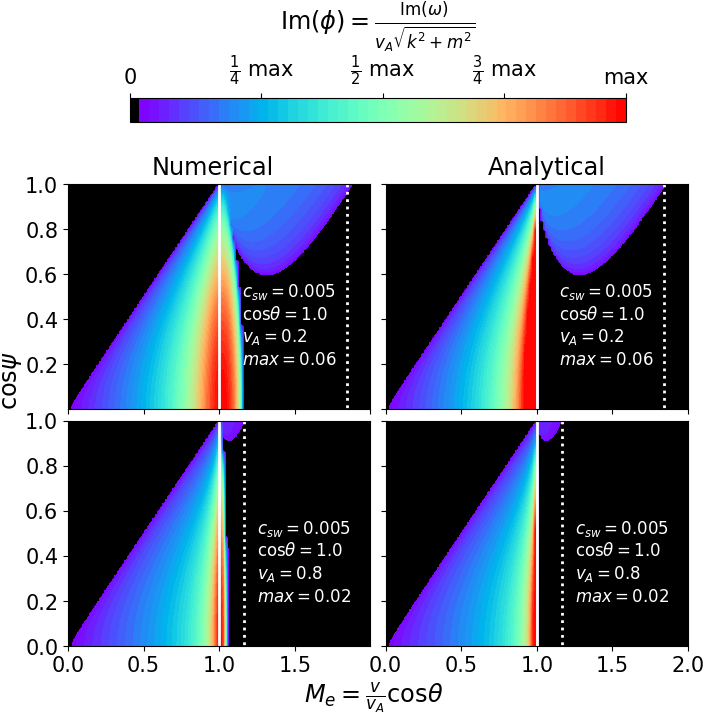}
\caption{Dependence of the instability growth rate $\text{Im}(\phi)$ on $\psi$ and $M_e$, for two choices of $\vatot$ as indicated in the plots. We fix $\cos\theta=1$. See the caption of Fig.~\ref{fig:numerical_vs_analytical_costheta} for further details.}
	\label{fig:numerical_vs_analytical_cospsi}
\end{figure}

In Fig.~\ref{fig:numerical_vs_analytical_costheta} and Fig.~\ref{fig:numerical_vs_analytical_cospsi}, we compare the numerical solution (left column) of  Eq.~(\ref{eq:interface_DR_param}) with our analytical approximation (right column). We fix $c_{sw}=0.005$ and consider $\vatot=0.2$ and 0.8, so the assumption $c_{sw}/\vatot\ll1$ of our analytical approximation is well satisfied. The analytical solution for $\text{Im}(\phi)$ displayed in the figures only employs the first order terms (as discussed above, we also use the second order terms to check the Sommerfeld constraint), yet it provides an excellent approximation of the numerical results, apart from $M_e=1$. For $M_e=1$, the first-order term $\Lambda_{\pm}$ of our analytical approximation diverges. We discuss below this special case.


Our analytical approximation allows to determine the range of $M_e$ where the system is unstable. If $\lambda$ in \eqq{approx_phi_lambda} is imaginary, then also $\Lambda_\pm$ has nonzero imaginary part. We then find the values of $M_e$ that satisfy $\lambda^2=0$ and obtain the following unstable bounds: for $M_e<1$,
\begin{align}
    \cos\psi < M_e < \min\left(\frac{\cos\theta}{\vatot},1\right) ~,\label{eq:bound1}
\end{align}
whereas for $M_e>1$
\begin{align}
    \sqrt{\frac{\nu_1 - \nu_2}{2+2\vatot^2\gammaw^2}} <M_e < \min\left(\frac{\cos\theta}{\vatot}, \sqrt{\frac{\nu_1 + \nu_2}{2+2\vatot^2\gammaw^2}}\right)~, \label{eq:bound2}
\end{align}
where
\begin{eqnarray}
\nu_1 & = & 2\cos^2\psi + (1+\vatot^2)\gammaw^2\\
\nu_2 & = & \sqrt{(1-\vatot^2)^2\gammaw^4-4(1-\cos^2\psi)(1-\vatot^2\cos^2\psi)\gammaw^2}\nonumber
\end{eqnarray}
Note that the condition $M_e<\cos\theta/\vatot$ is equivalent to the obvious requirement $v<1$. {\bf The condition $M_e>\cos\psi$ in \eqq{bound1} can be equivalently cast as $v\cos\theta>v_{\rm A}\cos\psi$, which has a simple interpretation. The system is unstable if the projection of the shear velocity onto the direction of $\bf{q}_\parallel$ (which we defined as the projection of the wavevector $\bf q$ on the $x-z$ plane, see Fig.~\ref{fig:schematic_setup}) is larger than the projection of the Alfv\'en speed onto the same direction. In other words, the shear  is able to overcome magnetic tension.}

\eqq{bound1} and \eqq{bound2} fully characterize the instability boundaries in Fig.~\ref{fig:numerical_vs_analytical_costheta} and Fig.~\ref{fig:numerical_vs_analytical_cospsi}. In particular, the vertical white dotted lines in the figures illustrate the upper bound in \eqq{bound2} for the special case $\cos\psi=1$, which yields
\begin{align} \label{eq:bounds_for_psi_1}
 1< M_e < \min\left(\frac{\cos\theta}{\vatot},\sqrt{\frac{1+\Gamma_w^2}{1+\vatot^2 \Gamma_w^2}}\right) \quad \text{for  } \cos\psi=1~.
\end{align}
It follows that the unstable range at $M_e>1$ shrinks for $\vatot\rightarrow 1$, but never disappears as long as $\vatot<1$.

\subsection{The special case $M_e$=$1$}
In the case $M_e=1$, our analytical approximation diverges.
The singular case  $M_e=1$ can be solved by expanding $\phi$ with a Puiseux series \citep{wall_2004, reference.wolfram_2022_asymptoticsolve}. Among the six approximate solutions of $\phi$ at $M_e$=$1$, the only unstable one is
\begin{align} \label{eq:pui1}
    \phi_{(M_e=1)} =&1 + (-1)^{2/3}(2\xi)^{1/3}\epsilon^{2/3},
\end{align}
where
\begin{align} \label{eq:pui2}
    \xi = \frac{(\cos^2 \psi -1)^2(\cos^2\theta-\vatot^2)}{\Gamma_w^2\cos^2\theta }~.
\end{align}
In Appendix \ref{app1} we demonstrate that this analytical approximation for the special case $M_e=1$ is in good agreement with the numerical solution.

\eqq{pui1} allows us to identify the range of $M_e$ (near unity) where the diverging growth rate in \eqq{approx1} should rather be replaced by \eqq{pui1}. By equating the imaginary parts of  $\phi_{(M_e<1)}$ in \eqq{approx1}  and $\phi_{(M_e=1)}$ in \eqq{pui1}, and solving for $M_e$, we can obtain the upper bound $M_e^*$ for \eqq{approx1} such that $\phi_{(M_e<1)} \leq\phi_{(M_e=1)}$ for $M_e\in [0,M_e^*]$. We expect $M_e^*$ to be close to unity, so we assume $M_e=1$ in $\mu$ and $\lambda$ for $\Lambda_+$ of \eqq{approx1}. The resulting expression for $M_e^*$ can then be written as
\begin{align}
    M_e^* = \sqrt{1-8\cdot3^{-1}(2\xi)^{1/3}\epsilon^{2/3}}~,
\end{align}
where we require $\epsilon<3^{3/2}2^{1/5}\xi^{-1/2}$ for real $M_e^*$.

\subsection{Maximum growth rate}
\begin{figure}[!h]
	\centering
	\includegraphics[width=0.5\textwidth]{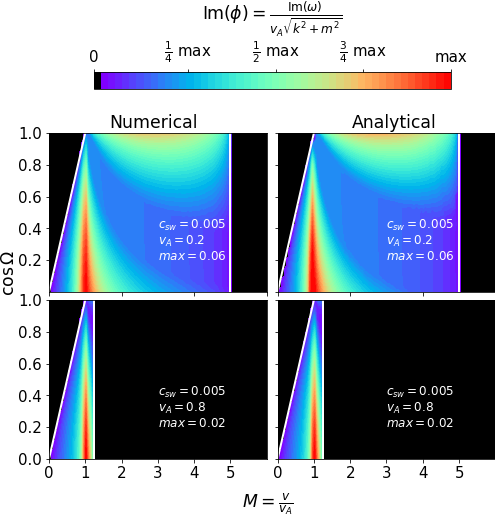}
 
\caption{{Dependence of the maximum instability growth rate $\text{Im}(\phi)$ on $\cos\Omega$ and $M\equiv v/\vatot$, for two choices of $\vatot$, as indicated in the plots. The maximum value of $\text{Im}(\phi)$ is taken across all values of $\cos \theta \in [0,1]$ for each $(M,\cos\Omega)$ pair. The left and right columns represent the numerical and analytical solutions, respectively. In all the panels, $\text{Im}(\phi)$ is then normalized to its maximum value, which is quoted in the panels themselves. The white lines indicate $M=\cos\Omega$ and $M=1/v_{\rm A}$.}}
\label{fig:numerical_vs_analytical_max_theta}
\end{figure}

The results presented so far retain the explicit dependence on the angle $\theta$ between the projected wavevector $\mathbf{q}_\parallel$ and the flow velocity $\mathbf{v}$, and on the angle $\psi$ between $\mathbf{q}_\parallel$ and the magnetic field $\mathbf{B}$ (see Fig.~\ref{fig:schematic_setup}). In practice, for a given Mach number $M=v/v_{\rm A}$ and a fixed magnetic field orientation (e.g., with respect to the shear direction), one can determine the maximum growth rate, irrespective of the specific value of $\theta$ at which it is attained. This is presented in \figg{numerical_vs_analytical_max_theta}, where we show the peak growth rate as a function of $M$ and $\cos\Omega$, where we define
\begin{equation}
\cos\Omega=\frac{\mathbf{v} \cdot \mathbf{B}}{|\mathbf{v}||\mathbf{B}|}
\end{equation}

The plot shows that, for most magnetic field orientations, the peak growth rate is achieved at $M\sim 1$. The exception is the case of fields nearly aligned with the shear velocity, where magnetic tension pushes the unstable region to higher $M$. The region of stability in the upper left corner is delimited by $M=\cos\Omega$ (white line), which comes from the instability condition $M_e>\cos\psi$ in \eqq{bound1}. The range of unstable Mach numbers extends up to $M<1/v_{\rm A}$ (vertical white line), which simply corresponds to the requirement $v<1$.

\section{Comparison to the hydrodynamic case}
When the unstable mode propagates perpendicularly to the magnetic field ($\cos\psi=0$), we expect magnetic tension to have no effect, and the solution should resemble the hydrodynamic asymmetric case discussed by \citet{blandford_pringle_1976}. We demonstrate this by
 choosing a different parameterization in Eq.~(\ref{eq:interface_DR_param}), similar to the one of Eq.~(2) in \citet{blandford_pringle_1976}, i.e. 
\begin{align}
    &\epsilon'=\frac{1}{\epsilon}=\frac{\vatot}{c_{sw}},\quad \phi'=\frac{\phi}{\epsilon}=\frac{\omega}{c_{sw}\sqrt{k^2+m^2}},\nonumber \\
    &\delta'
    =\frac{w_{0w}}{w_{0j}^*}\frac{c_{sw}^2}{\vatot^2}, \quad \eta'=\vatot\epsilon=c_{sw}, \nonumber \\
    &M'=\frac{M \cos\theta}{\epsilon}=\frac{v}{c_{sw}}\frac{k}{\sqrt{k^2+m^2}},
    \label{eq:primed}
\end{align}
where $w_{0j}^*$ is the total enthalpy of the jet, namely the sum of the gas enthalpy $w_{0j}$ and the magnetic enthalpy:
\begin{align}
    w_{0j}^* = B_{0x}^2+B_{0z}^2 + w_{0j} = \frac{w_{0j}}{1-\vatot^2}~.
\end{align}
Then the dispersion relation Eq.~(\ref{eq:interface_DR_param}) can be equivalently written as
\begin{align}
    &(\phi'^2-\cos^2\psi)^2[\gamma^2(1-\eta'^2)(\phi'-M')^2+\eta'^2\phi'^2-1] \nonumber \\
    & \quad = \gamma^4\delta'^2(\phi'-M')^4(\phi'^2-\epsilon'^2)\epsilon'^2,
\end{align}
which, by setting $\cos\psi=0$, is exactly the same as Eq.~(1) in \citet{blandford_pringle_1976}, where both the jet and the wind were assumed to be unmagnetized. We conclude that, even though our jet is magnetized, in the case $\cos\psi=0$ the instability behaves similarly to the case of a hydrodynamic jet. Here, the magnetic field provides pressure, but not tension. 

\section{Discussion and conclusions}
We have studied the linear stability properties of the KHI for relativistic, asymmetric, magnetized flows, with focus on conditions appropriate for the interface between a magnetized relativistic jet and a gas-pressure-dominated wind. We derive the most general form of the dispersion relation and provide an analytical approximation of its solution for $\epsilon=c_{sw}/\vatot\ll1$. The stability properties are chiefly determined by the angle $\psi$ between the jet magnetic field and the  wavevector projection onto the jet/wind interface. For $\psi=\pi/2$, magnetic tension plays no role, and our solution resembles the one of a gas-pressure dominated jet. Here, only sub-Alfv\'enic jets are unstable ($0<M_e\equiv(v/\vatot)\cos\theta<1$, as long as $v<1$). For $\psi=0$, the velocity shear needs to overcome the magnetic tension, and only super-Alfv\'enic jets are unstable ($1<M_e<\sqrt{(1+\Gamma_w^2)/(1+\vatot^2\Gamma_w^2)}$). At zeroth order in $\epsilon$, the phase speed of unstable modes is $\omega/k=v$ in the jet frame, i.e., they are purely growing in the wind frame.





Our analytical results are valuable for both theoretical and observational studies. They can be easily  incorporated into global MHD simulations of jet launching and propagation, to identify KH-unstable surfaces \citep{chatterjee2020,sironi_21,wong2021}. On the observational side, claims have been made that the KHI is observed along Active Galactic Nuclei (AGN) jets, based on the geometry of the outflow \citep{lobanov2001,issaoun2022}. Our formulae can place this claim on solid grounds, if estimates of the field strength and orientation and of the flow velocities are available. Besides AGNs, our results have implications for other jetted sources such as, but not limited to, gamma-ray bursts, tidal disruption events, X-ray binaries, and pulsar wind nebulae.

We conclude with a few caveats. First, the plane-parallel approach we employed is applicable only if the jet/wind interface is much narrower than the jet radius (for studies of surface instabilities in force-free cylindrical jets see, e.g., 
\citealt{Bodo_Mamtsashvili_Rossi_2013,Sobacchi_18,Bodo_Mamtsashvili_Rossi_2016, Bodo_Mamtsashvili_Rossi_2019}). Secondly, our local description implicitly assumes that the flow properties do not change as the KHI grows.
Third, we have assumed the jet plasma to be cold, and the surrounding medium to be unmagnetized. These assumptions will be relaxed in a future work.



\begin{acknowledgements}
We are grateful to R. Narayan for many inspiring discussions and collaboration on this topic. We would like to thank G. Bodo for many
useful discussions and suggestions. L.S. acknowledges support from the Cottrell Scholars Award and the DoE Early Career Award DE-SC0023015. L.S and J.D. acknowledge support from NSF AST-2108201, NSF PHY-1903412 and NSF PHY-2206609. J.D. is supported by a Joint Columbia University/Flatiron Research Fellowship, research at the Flatiron Institute is supported by the Simons Foundation.
\end{acknowledgements}


\appendix
\section{Analytical approximation for $M_{e}=1$}\label{app1}
\begin{figure}[!ht]
	\centering
	\includegraphics[width=0.5\textwidth]{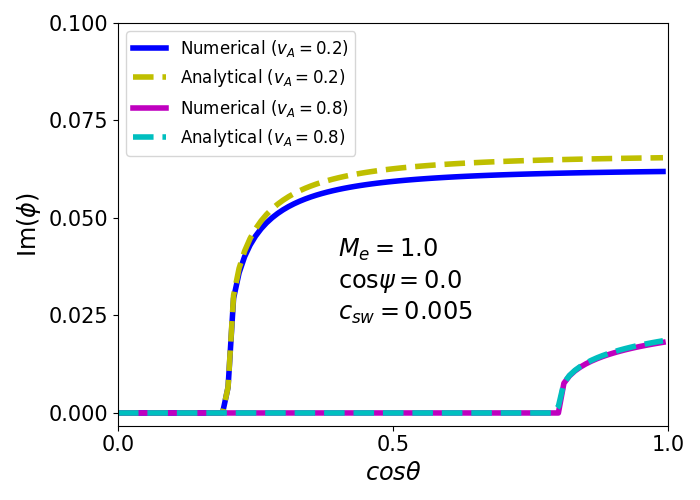}
\caption{Dependence of the instability growth rate Im$(\phi)$ on $\cos\theta$ for two choices of $\vatot$ and a fixed value of $\cos\psi=0$ in the singular case  $M_e=1$. Solid lines represent the numerical solutions while  dashed lines represent the analytical solutions obtained by Puiseux series expansion in the main text.
}\label{fig:numerical_vs_analytical_puiseux_theta}
\end{figure}

\begin{figure}[!ht]
	\centering
	\includegraphics[width=0.5\textwidth]{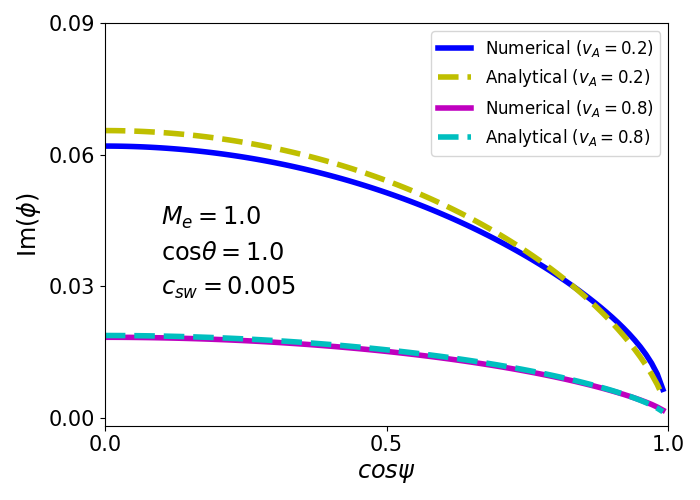}
\caption{Dependence of the instability growth rate Im$(\phi)$ on $\cos\psi$ for two choices of $\vatot$ and  a fixed value of $\cos\theta = 1$ in the singular case  $M_e=1$. Solid lines represent the numerical solutions while  dashed lines represent the analytical solutions obtained by Puiseux series expansion in the main text.
}\label{fig:numerical_vs_analytical_puiseux_psi}
\end{figure}

For the singular case $M_e=1$, our analytical solutions take the form of the first order Puiseux series. Here we compare the analytical and numerical solutions. In Fig.~\ref{fig:numerical_vs_analytical_puiseux_theta} and Fig.~\ref{fig:numerical_vs_analytical_puiseux_psi}, we plot the instability growth rate for $M_e=1$, comparing analytical and numerical solutions. We choose the same parameters as in the figures of the main paper, namely $c_{sw}=0.005$, and $\vatot=0.2$ or $0.8$. We fix $\cos\psi=0$ for Fig.~\ref{fig:numerical_vs_analytical_puiseux_theta} and $\cos\theta=1$ for Fig.~\ref{fig:numerical_vs_analytical_puiseux_psi}. We use solid and dashed lines to represent numerical and analytical solutions, respectively. The figures show that our analytical solutions in Puiseux series provide a good approximation to the numerical ones across the entire range of $\cos\theta$ (for Fig.~\ref{fig:numerical_vs_analytical_puiseux_theta}) and $\cos\psi$ (for Fig.~\ref{fig:numerical_vs_analytical_puiseux_psi}).

\section{The second order terms}\label{app2}
In the main body of the paper, we have looked for an analytical approximation of the form
$\phi \approx c_0 + c_1 \epsilon + c_2\epsilon^2$,
where $c_0, c_1$ and $c_2$ are constant with respect to $\epsilon$ and terms higher than $\epsilon^2$ are ignored. For the unstable solutions, we find that the first order term $\Lambda_{\pm}\epsilon$ generally provides a good approximation of the numerical solution. However, the second order term $\Sigma_\pm \epsilon^2$ is
required for identifying the physical solutions that satisfy the Sommerfeld condition. The explicit expression for $\Sigma_\pm$ is
\begin{equation}
        \Sigma_\pm =\frac{M_e \mu [(1 - M_e^2) (
      \cos^2\psi (1 - \vatot^2) + M_e^2 (1 + 3 \vatot^2)-2 - 2 M_e^4 \vatot^2)
\Gamma_w^2 +
   2 ( \cos^2\psi + M_e^2-2) (\mu^2 \pm
     \lambda)]}{\gamma^2(1-M_e^2)^2\Gamma_w^2\lambda}\
\end{equation}


\providecommand{\noopsort}[1]{}

\end{document}